\documentclass[floatfix, preprintnumbers, amsmath, amssymb, twocolumn, 10pt, showpacs]{revtex4}
\usepackage{dcolumn}
\usepackage{bm}
\usepackage{epsfig}
\usepackage{graphicx}
\usepackage{amsmath}
\usepackage{amssymb}
\begin{document}

\date{\today}
\preprint{Brown-HET-1456}
\title{On the Transfer of Metric Fluctuations when Extra Dimensions Bounce or Stabilize}

\author{Thorsten J. Battefeld$^{1)}$} \email[email: ]{battefeld@physics.brown.edu}
\author{Subodh P. Patil $^{1,2)}$} \email[email: ]{patil@het.brown.edu}
\author{Robert H. Brandenberger$^{2)}$} \email[email: ]{rhb@hep.physics.mcgill.edu}

\affiliation{1) Dept.of Physics, Brown University, Providence R.I. 02912, U.S.A.} 
\affiliation{2) Dept.of Physics, McGill University, Montr\'eal QC, H3A 2T8, Canada}

\begin{abstract}
In this report, we study within the context of general relativity with one extra 
dimension compactified either on a circle or an orbifold, how radion fluctuations 
interact with metric fluctuations in the three non-compact directions. The background 
is non-singular and can either describe an extra dimension on its way to stabilization, 
or immediately before and after a series of non-singular bounces. We find that the 
metric fluctuations transfer undisturbed through the bounces or through the transients 
of the pre-stabilization epoch. Our background is obtained by considering the effects 
of a gas of massless string modes in the context of a consistent 'massless background' 
(or low energy effective theory) limit of string theory. We discuss applications to 
various approaches to early universe cosmology, including the ekpyrotic/cyclic 
universe scenario and string gas cosmology.
\end{abstract}

\maketitle

\section{Introduction}

The idea that the universe is comprised of any number of extra dimensions, in addition 
to the three non-compact spatial dimensions that we observe, goes back all the way to 
the work of Kaluza and Klein in the 1920's \cite{klein,kaluza}, where extra dimensions 
were first proposed as a manner in which one could unify particle interactions with 
gravity. Since then, theories with extra dimensions have evolved from being a 
mathematical curiosity, albeit one with remarkable consequences for particle physics 
\cite{pope}, to a theoretical necessity with the advent of string theory. Since the 
famous anomalies cancellation calculation of Green and Schwarz \cite{gs}, the challenge 
has remained for cosmologists and phenomenologists alike to account for the presence 
of extra dimensions in physics. From a pessimist's point of view, one might view these 
extra dimensions as an unwanted surplus that has to be swept under the rug at presently 
accessible energies. In its most unimaginative form, one could view this as the gist 
of the moduli problem. From an optimist's point of view however, one could view the 
requirement of extra dimensions for the consistency of string theory as a new window 
of opportunity, through which one might be able to resolve some of the outstanding 
problems of particle physics and cosmology. 

\par

Taken as an essential part of the radically new picture of space-time that string 
theory offers us, extra dimensions, with strings and branes of various dimensions 
propagating through them, could offer us a potentially non-anthropic explanation 
for the observed dimensionality of space-time \cite{BV}. In addition to offering us 
potentially non-singular cosmologies \cite{GV,BV,TV}, they could very well be at the 
heart of the resolution of the dark matter problem 
\cite{Nusser:2004qu,Battefeld:2004xw,Gubser:2004du,Gubser:2004uh}. Recently, 
several alternative stringy cosmologies have emerged as possible successors to 
the standard big bang model of the universe, such as the pre-big bang model \cite{GV}, 
brane/string gas cosmology \cite{BV,TV,Brandenberger:2005fb,Brandenberger:2005nz,Battefeld:2005av}
, and the 
cyclic/ekpyrotic scenarios \cite{Khoury:2001wf,st1,st2,st3,st4,brown}, all of which 
have the ultimate aim of becoming complete and testable models of the early universe. 

\par

A central question that arises in determining whether or not these models of the 
early universe reproduce observations is, how metric fluctuations in the presence of 
extra dimensions evolve in the backgrounds that these cosmologies propose. The 
importance of understanding this question cannot be understated in the context of the 
abundance of available experimental data, against which we must compare our eventual 
predictions. For instance, in the ekpyrotic/cyclic scenario the effects of branes 
colliding in a 5-dimensional bulk are explored \footnote{Note that, as stressed in \cite{Pyro}, none of these
 alternatives at the present stage solve the homogeneity and flatness
 problems of standard big bang cosmology without invoking a period
 of inflation (or something which effectively acts as inflation).}

Key to the success of this program is being able 
to follow the evolution of metric fluctuations through the bounces of the extra 
dimension. The eventual goal of this is to be able to explain the observed 
inhomogeneities of the cosmic microwave background as being seeded by radion 
fluctuations generated through the motion of the branes towards each other, rather 
than through the quantum fluctuations of some putative inflaton field. 

\par

Similarly, in the context of the pre-big bang scenario, where an expanding and a 
contracting phase of the universe's evolution are naturally related to each other by 
the scale factor duality symmetry inherent to low energy effective string theory, 
a universe undergoing a big bang-big crunch transition is naturally implemented. 
In this context, the evolution of metric fluctuations through the bounce becomes a 
question of central importance in understanding the way the big bang phase of our 
universe's evolution happened the way it did. As it turns out, although at first 
seemingly unrelated, an associated and as yet unexamined issue is how any tentative 
solution of the moduli problem will modify the spectrum of 
the metric fluctuations which we observe. Given that the universe is likely to 
contain many extra dimensions whose shape and volume moduli are dynamical in
the early universe, one might expect that the dynamics of these moduli
fields will lead to a highly nontrivial modification of any initial 
pre-stabilization spectrum.  

\par

It is the goal of this report to show that, in the context of a 5-d universe where 
an extra dimension undergoes non-singular bounces, or is en route to stabilization, 
the final spectrum for the Bardeen potential corresponding to (long wavelength) 
fluctuations of the scale factor for the non-compact dimensions, is identical to the 
initial spectrum of the five-dimensional fluctuations (in a sense which will
be specified later). Specifically, if we had an initial scale invariant spectrum for 
the five-dimensional fluctuations, we would end up with a scale invariant spectrum 
for the Bardeen potential after the bounce, or after the transients of the 
stabilization had settled down. Furthermore, we show that the variable corresponding 
to metric fluctuations of the compact dimensions decays, which corresponds to the 
stability of the radion degree of freedom to fluctuations in our non-singular setup.

\par

One application of our work concerns the evolution of fluctuations in ekpyrotic/cyclic
type models. In the context of a four space-time dimensional effective field
theory toy model of this scenario, the dynamics of perturbations 
has been investigated in detail. The
initial analyses \cite{Khoury:2001wf,KOST2} yielded the result that a scale-invariant
spectrum before the bounce transfers to a scale-invariant spectrum after the
bounce. These analyses, however, were questioned in 
\cite{Lyth1,FB,Lyth2,Hwang,Tsujikawa,IAP}. A serious complicating factor turned
out to be the fact that the proposed background evolution was singular, thus
requiring the use of ``matching conditions'' (such as those derived in \cite{DM}
in the context of an expanding cosmology undergoing a sudden phase transition)
to compute the post-bounce spectrum of fluctuations. As discussed in \cite{Durrer},
there is a very sensitive dependence of the final result on the choice of the
matching surface. Independently, there has been recent work on the
evolution of fluctuations through a nonsingular bounce in four space-time
dimensional cosmologies, in which the bounce is constructed by adding extra
terms to the standard Lagrangian. These analyses 
\cite{Paris1,Paris2,TFB,FF03,Wands,Bozza:2005wn,Bozza:2005xs,Battefeld:2005cj,Geshnizjani:2005hc} 
yield results showing 
a sensitive dependence on the nature of the bounce. The bottom line of
this work is that a correct analysis of fluctuations in the ekpyrotic/cyclic
scenario needs to be done in a five space-time dimensional context, a
context in which the nature of the bounce is unambiguous. A first important
step in this direction was taken in \cite{Tolley}, confirming the result that
a scale-invariant spectrum passes through the bounce (in the five-dimensional
context, the bounce means that the radius of the extra spatial dimension bounces,
not that the four space-time dimensional scale factor bounces) without change in the
spectral index. However, the analysis of \cite{Tolley} was done in the
context of a singular background and assuming specific matching conditions
for fluctuations applied at a point when the perturbations in fact blow up.
Thus, the results are open to doubt. In this work, we study the transfer
of fluctuations through a cosmology in which two boundary branes approach
each other and bounce without encountering a singularity (see also \cite{brown}
for previous work done in the context of a particular nonsingular ekpyrotic-type
bounce proposed in \cite{Rasanen}).

We begin by introducing a non-singular bouncing model of a 5-dimensional universe 
where one dimension is compactified on a circle or an orbifold. As we shall see 
shortly in detail, our non-singular background is obtained by considering the effects 
of gas composed of massless string modes on the dynamics of space-time, in the context of a 
consistent low energy effective theory limit, or `massless background',  of string 
theory. We then set up the framework for studying cosmological perturbations in this 
model and  study how these transfer through the various bounces that the extra 
dimension undertakes. Although our setup is seemingly specific to string gas cosmology, 
the essence of our framework is that we have a non-singular bounce/stabilization 
mechanism that is affected by degrees of freedom that become massless at a certain 
point. Hence, the hope is that the results obtained here can be generalized to other 
settings, an issue we will discuss in detail when we consider applications to 
different approaches to stringy cosmology. We now commence our paper with a few 
preliminaries.  

\section{The background -- $\mathbb R^4 \times S^1$ \label{secII}}

Consider a five dimensional space-time with the topology of $\mathbb R^4 \times S^1$, 
described by the metric \footnote{The metric can always be cast into this form, such that $t$ corresponds to cosmic time.}
\begin{equation}
\label{metric}
g_{AB} = diag(-1,a^2(t),a^2(t),a^2(t),b^2(t))\,,
\end{equation}
from which we derive the following components of the Einstein tensor:
\newcommand{\hub}[1]{\frac{\stackrel{.}{{#1}}}{{#1}}}
\newcommand{\hubd}[1]{\frac{\stackrel{..}{{#1}}}{{#1}}}
\begin{eqnarray}
\label{00} G^t_{\,\,t} &=& -3\hub{a} \Bigl( \hub{a} + \hub{b} \Bigr)\,, \\ 
\label{jj} G^{x_i}_{\,\,x_j} &=& -\delta^i_j \Bigl[ 2\hubd{a} + \hubd{b} + \Bigl( \hub{a} \Bigr)^2 +2\hub{b}\hub{a} \Bigr]\,, \\ 
\label{55} G^y_{\,\,y} &=& -3 \Bigl[ \hubd{a} + \Bigl( \hub{a} \Bigr)^2 \Bigr]\,.
\end{eqnarray}
Here, the indices $i$ and $j$ run over the three large spatial dimensions and $y$ denotes the extra dimension. The form of 
the energy-momentum tensor which will couple to the Einstein tensor is given by
\begin{equation}
\label{emt}
T^A_{\,\,B} = diag(-\rho,p,p,p,r)\,. 
\end{equation} 

We can recast the Einstein equations $G^A_{\,\,B} = 8\pi G T^A_{\,\,B}$, where $G$ is the 
five-dimensional Newton's constant, in the form:
\begin{eqnarray}
\label{a}
\hubd{a} &+& H(2H + \mathcal H) + \frac{8\pi G}{3}[r - \rho] = 0\,,\\
\label{b}
\ddot{b} &+& 3H\dot{b} + 8\pi G~b[p - \frac{2r}{3} - \frac{\rho}{3}] = 0\,,\\
\label{hc}
\rho &=& \frac{3}{8\pi G}H(H + \mathcal H)\,,
\end{eqnarray}
where $H=\dot{a}/a$ and $\mathcal{H}=\dot{b}/b$. The energy-momentum tensor for a 
string gas in this toroidally compactified background was derived in \cite{sp1} to 
which we refer the reader if any of what follows is unfamiliar. It was found to be
\begin{eqnarray}
\label{tt}
\rho_{n,w} &=& \frac{\mu_{0,n,w}}{\epsilon_{n,w}\sqrt{-g}} \epsilon^2_{n,w}\,,\\
\label{ii}
p_{n,w} &=& \frac{\mu_{0,n,w}}{\epsilon_{n,w}\sqrt{-g}}\frac{p^2_{n.c.}}{3}\,,\\
\label{aa}
r_{n,w} &=& \frac{\mu_{0,n,w}}{\epsilon_{n,w}\sqrt{-g}\alpha'}\Bigl( \frac{n^2}{\tilde{b}^2} - w^2\tilde{b}^2 \Bigr)\,,
\end{eqnarray}    
where $g$ denotes the determinant of the metric, $\tilde{b}:=b/\sqrt{\alpha^{\prime}}$ (with $2\pi\alpha^\prime$ being the inverse string tension) and the subscripts $n,w$ refer to the 
momentum and winding quantum numbers of a 
closed string along the 5th dimension respectively, on which all of the subscripted 
quantities will depend. The density $\mu_0$ is the number density of the string gas 
with the metric dependence factored out ($\mu(t) = \mu_0/\sqrt{-g}$), 
while $\epsilon_{n,w}$ is the energy of a single closed string in this background
\begin{equation}
\label{e}
\epsilon_{n,w} = \sqrt{p_{n.c.}^2 + \left(\frac{n}{b} + \frac{wb}{\alpha'}\right)^2 + \frac{4}{\alpha'}(N-1)}\,,
\end{equation}  
where $p_{n.c.}$ denotes the center of mass momentum along the three 
non-compact directions, and $N$ is the number of right-moving string vibrational
modes. For reasons made clear in \cite{sp1} and \cite{sp2}, 
string states that are massless at special symmetry points (i.e. the self-dual radius) 
should play a very distinguished role in any string gas cosmology (see also 
\cite{Watson:2003gf, Kofman:2004yc, Watson:2004aq}). 
There are several reasons for this, among which are prominently the desire to
obtain a viable late time phenomenology and a robust stabilization mechanism for 
the radion. However, the main motivation is that the low energy effective theory 
limits of string theory (such as general relativity and dilaton gravity) are also 
consistent backgrounds on which we can study the propagation of massless 
(and only massless) strings. For these reasons we do not further justify the focus
on massless string modes, and refer the reader to \cite{sp1,sp2,sp3} for more details. For a general review of string gas cosmology, we refer the interested reader to \cite{Brandenberger:2005fb}\cite{Brandenberger:2005nz}\cite{Battefeld:2005av}.  

We wish to emphasize that the framework within which we chose to work  assumes nothing other than the fact that nature is described by a string theory at high energies, and that the degrees of freedom that are likeliest to be excited (namely massless string modes) will be excited. The two particular geometries that we consider for the extra dimensions (i.e. a toroidal geometry in this section and an orbifold geometry in the next section), are necessitated by the fact that they are the unique compact backgrounds on which one stays within the approximation of the low energy effective limit of string theory, when considering extra dimensions that are similar in size to the string scale. This is because in general, were the metric to depend on the extra dimension (as is the case for warped extra dimensions), one will introduce curvatures that approach the string scale when the extra dimensions themselves approach the string scale, and hence the low energy approximation will break down. We discuss this point further in the next section.

\par
The particular states we are interested in are those for which $n = -w = \pm 1, N = 1$. 
That these states are massless at the self-dual radius ($b = \sqrt{\alpha'}$) is 
easily checked from (\ref{e}). These states imply that (\ref{b}) becomes
\begin{equation}
\label{blah}
\ddot{\tilde{b}}+3\frac{\dot{a}}{a}\dot{\tilde{b}}+\frac{8\pi G\mu_0}{a^3\alpha^{\prime}}\frac{-\frac{1}{\tilde{b}^2}+\frac{\tilde{b}^2}{3}+\frac{2}{3}}{\sqrt{\alpha^{\prime}p_{n.c.}^2+\left(\frac{1}{\tilde{b}}-\tilde{b}\right)^2}} = 0\,.
\end{equation} 
We will see that this background 
stabilizes around the self-dual radius ($\tilde b = 1$), whilst the non-compact 
directions persistently expand as they would in a radiation dominated universe. 
Before we get to this, however, we wish to discuss a generalization of this 
background to the situation where the extra dimension is compactified on an orbifold.

\subsection{Extension to $\mathbb R^4 \times S^1/\mathbb Z_2$ \label{secIII}}

The consistency of the framework that we propose, namely that we remain within the 
low energy effective theory (or massless background) limit of string theory, requires 
the background we consider to satisfy the condition
\begin{equation}
\label{approx}
R[g] \ll \frac{1}{\alpha'}\,. 
\end{equation} 
That is, the Ricci scalar should be considerably bounded from above by the string 
tension. This requirement translates into the statement that the metric should not 
change very much on the string scale, which ensures that this background remains a 
consistent background for the propagation of massless string modes. Conversely, were 
one to consider backgrounds which do not satisfy (\ref{approx}), not only would the 
approximations made in deriving the low energy limit of string theory break down, one 
would also expect massive modes to be created \cite{pol}. For our purposes, the 
requirement of (\ref{approx}) means that were we to consider compactifications of the 
extra dimension on scales comparable to the self-dual radius, then there should be 
absolutely no dependence of the metric along the compactified direction. This is a 
remarkable feature of using a toroidal compactification: one can in fact study string 
scale processes without invalidating the approximations inherent in the low energy 
limit of string theory. It was shown in \cite{sp2} and \cite{sp3} that the background 
we are about to derive, does in fact satisfy (\ref{approx}) throughout its dynamics. 

\par

Returning to the problem at hand, we see that since we are not allowed to consider 
metric dependencies along the extra dimension (by homogeneity and isotropy of the 
non-compact dimensions, the metric can then only depend on time), the components of 
the Einstein tensor (\ref{00})-(\ref{55}) are unchanged after orbifolding the extra 
dimension. The only place where orbifolding might make a difference is in the energy
momentum tensor of the string gas. However, because the massless states that we have 
focused on are in the so-called untwisted sector \cite{pol} (in general, twisted 
states are localized at the orbifold fixed points), it turns out that there is no 
difference in the energy-momentum tensor either (\ref{tt})-(\ref{aa}). This is a 
consequence of the `inheritance' principle of orbifold theories \cite{pol}, and 
permits us to use the framework just presented in either situation, provided we only 
use the modes that we have indicated and we do not consider any variations in any 
metric quantities along the extra dimension. However, there is a caveat to this in 
that two orbifold fixed planes are present, which might have matter confined to live 
on them. The requirement not to induce any metric variations along the extra dimension 
translates into, via the Israel junction conditions \cite{Israel:1966rt,deff}, the 
condition (among others)
\begin{equation}
\label{ijc}
\left[\frac{d a}{dy}\right] \propto 8\pi G \rho_{brane}, 
\end{equation}
where $[da/dy]$ is the jump of the derivative of the scale factor $a$ along 
the extra dimension, evaluated around either of the orbifold fixed planes. Hence, we 
see that if we require there to be no dependence of the metric along the extra 
dimensions, then any matter localized on the branes must be sufficiently dilute to 
render the branes to behave as test branes. Bearing all of this in mind, we can now 
proceed to derive the background solution within our framework.

\section{Background solution in 5D}

\subsection{The three large dimensions}

We first note from (\ref{b}) that any form of matter which satisfies a radiative 
equation of state ($p=\rho/3$, $r=0$) for all times (as opposed to our wound string 
states which become massless, hence radiative,  only at the self-dual radius) drops 
out of the driving term in the equation of motion for $b$ (\ref{b}). Hence, we can 
safely consider a situation where we have a radiation gas in addition to our gas of 
wound strings, which dominates the evolution of the Hubble factor $H$ for the 
non-compact dimensions through (\ref{hc}) after the extra dimension has stabilized, 
or is close to stabilization. In fact, massless unwound closed string states 
(gravitons: $n=w=0$, $N=1$) can provide exactly such a bath. Henceforth, 
it is natural to include such a bath in our setup.  We do not consider this matter 
further, simply taking as a given in what follows that the scale factor $a$ for the 
non-compact dimensions expands at the background level as it would in a radiation 
dominated universe, that is
\begin{equation}
\label{ast}
a(t) \propto t^{1/2}\,.
\end{equation}

\subsection{The extra dimension}

\begin{figure}[tb]
  \includegraphics[width=\columnwidth]{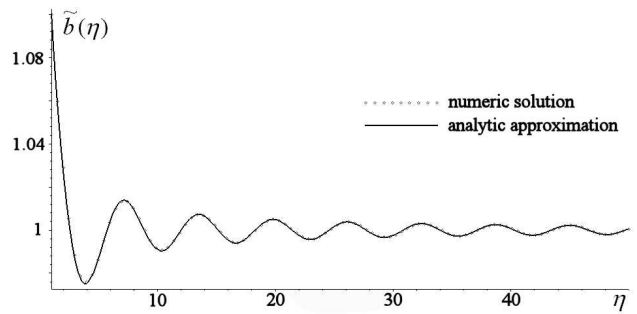}
   \caption{\label{fig:b_a_n} The analytic approximation (\ref{analyticb}) (solid line) is compared with the numeric solution of (\ref{eom}) (circles). }
\end{figure}

Let us go back to the equation of motion (\ref{blah}) for $\tilde{b}$, which reads
\begin{eqnarray}
0=\ddot{\tilde{b}}+3\frac{\dot{a}}{a}\dot{\tilde{b}}+\frac{8\pi G\mu_0}{a^3\alpha^{\prime}}\frac{-\frac{1}{\tilde{b}^2}+\frac{\tilde{b}^2}{3}+\frac{2}{3}}{\sqrt{\alpha^{\prime}p_{n.c.}^2+\left(\frac{1}{\tilde{b}}-\tilde{b}\right)^2}}\,,\label{eomsubodh}
\end{eqnarray}
where $p_{n.c.}=:q/a$ with $q=\mbox{const}$ is the
center of mass momentum along the three 
large dimensions. Introducing the characteristic time scale
\begin{eqnarray}
t_0^{-1}:=\sqrt{\frac{64\pi G\mu_0}{3\alpha^{\prime 3/2}a_0^2q}}
\end{eqnarray}
and defining the dimensionless time variable
\begin{eqnarray}
\eta:=2\sqrt{\frac{t}{t_0}}\,,\label{defeta}
\end{eqnarray}
such that $a=a_0\eta$, we can simplify (\ref{eomsubodh}) to
\begin{eqnarray}
\tilde{b}^{\prime\prime}+\frac{2}{\eta}\tilde{b}^{\prime}+\frac{3}{8}\frac{-\frac{1}{\tilde{b}^2}+\frac{\tilde{b}^2}{3}+\frac{2}{3}}{\sqrt{1+\frac{\eta^2}{4}l^2\left(\frac{1}{\tilde{b}}-\tilde{b}\right)^2}}=0\,, \label{eom}
\end{eqnarray}
where a prime denotes a derivative with respect to $\eta$  and we also introduced the 
free dimensionless parameter
\begin{eqnarray}
l:=\frac{a_0}{\sqrt{\alpha^{\prime}}q}\,.
\end{eqnarray}
This equation can easily be integrated numerically, but it will turn out to be useful 
to have a simple analytic approximation.

If we are close to the self-dual radius, that is $\tilde{b}=1+\varepsilon$ with 
$\varepsilon\ll1$,  we can expand the driving term in (\ref{eom}) so that
\begin{eqnarray}
\varepsilon^{\prime\prime}+\frac{2}{\eta}\varepsilon^{\prime}+\varepsilon=0
\end{eqnarray}
results. The general solution is given by
\begin{eqnarray}
\varepsilon(\eta)=\frac{1}{\eta}\left(A\sin(\eta)+B\cos(\eta)\right)\,, \label{analyticb}
\end{eqnarray}
with $A$ and $B$ constants. This solution may either describe the stabilization of the 
extra dimension or a series of non-singular bounces \footnote{Note that this series of bounces differs from the one in e.g. the cyclic scenario, since $\epsilon$ is small.}. See Fig.\ref{fig:b_a_n} for a comparison of the numeric solution of (\ref{eom}) with the analytic one. Since they are barely discernable we will use the analytic 
approximation in the following and switch freely between bounce/stabilization-language.

\section{Scalar perturbations in 5D}

We will first perturb the metric, focusing only on scalar metric perturbations.
The most general scalar perturbation in ``generalized'' longitudinal gauge 
\cite{Mukhanov:1990me,vandeBruck:2000ju,brown}
can be characterized by four scalar functions, $\Phi ,\Psi ,W$ and $\Gamma$
. These functions can be viewed as a basis of gauge
invariant variables. In this gauge, the metric is given by
\begin{eqnarray}
\nonumber ds^2&=&-(1-2\Phi (t,{\bf x}))\,dt^2
+a(t)^2(1-2\Psi(t,{\bf x})) d {\bf x}^2 \\
 &&+b(t)^2(1+2\Gamma(t,{\bf x}))\,dy^2
-2W(t,{\bf x}) \,dt\,dy \,,\label{perturbedmetrik}
\end{eqnarray}
where the signs in front of the perturbations are a mere convention \footnote{In our notation $\Phi=-\Psi$ would be called the Newtonian potential in a four dimensional description.}.  For reasons 
discussed in sections \ref{secII} and \ref{secIII}, we do not consider any $y$ 
dependence, that is we assume homogeneity in the extra dimension (see also \cite{Battefeld:2005av} for a discussion of this and other assumptions of SGC). The corresponding 
Einstein tensor can be computed to be \cite{vandeBruck:2000ju,brown} (in the
following $\Delta$ denotes the Laplacian of the three large spatial dimensions):
\begin{eqnarray}
\delta G^{x_i}_{\,\,x_j}&=&\frac{1}{a^2}\partial _{x_i}\partial _{x_j}
\left[\Psi +\Phi -\Gamma\right]\, \,,\,\,\ i \neq j,\label{G^i_j;ineqj}\\
\delta G^y_{\,\,t}&=&
\left[\frac{3}{b^2}\left(
\frac{\ddot{a}}{a}-\frac{\dot{a}\dot{b}}{ab}\right)
+\frac{1}{2b^2a^2}\triangle\right]W \,,\label{einsteinyt}\\
\delta G^{y}_{\,\,x_i}&=&\partial _{x_i}\Bigg[\frac{1}{2b^2}\left(\frac{\dot{b}}{b}+\frac{\dot{a}}{a}
+\partial _t\right)W\Bigg]\label{einsteinyxi}
\,,
\end{eqnarray}

\begin{widetext}
\begin{eqnarray}
\delta G^{x_i}_{\,\,x_i}&=&\frac{1}{a^2}
 \left[\partial _{x_i}\partial _{x_i}-\triangle\right]\left(
-\Gamma +\Psi +\Phi \right)
+2\left(\frac{\dot{b}}{b}
+3\frac{\dot{a}}{a}+\partial _t\right)
\partial _t\Psi \label{einsteinxixi}\\
\nonumber &&-2\left(\frac{\ddot{b}}{b}
+2\frac{\dot{a}\dot{b}}{ab}
+\frac{\dot{a}^2}{a^2}+2\frac{\ddot{a}}{a}
+\frac{\dot{b}}{2b}\partial _t
+\frac{\dot{a}}{a}\partial _t\right)
\Phi-\left(2\frac{\dot{a}}{a}
+2\frac{\dot{b}}{b}+\partial _t
\right)
\partial _t\Gamma \,,\\
\delta G^{y}_{\,\,y}&=&\left[-\frac{2}{a^2}\triangle +3\left(\partial _t
+4\frac{\dot{a}}{a}\right)\partial _t
\right]\Psi
+\left[-\frac{1}{a^2}\triangle
-6\left(\frac{\dot{a}^2}{a^2}
+\frac{\ddot{a}}{a}+\frac{\dot{a}}{2a}\partial _t\right)
\right]\Phi
 \,, \label{einsteinyy}\\
\delta G^t_{\,\,t}&=&\left[ 3\left(\frac{\dot{b}}{b}\partial _t
+2\frac{\dot{a}}{a}\partial _t\right)-\frac{2}{a^2}\triangle\right]\Psi
-6\left(\frac{\dot{a}\dot{b}}{ab}+\frac{\dot{a}^2}{a^2}\right)
\Phi
-\left(\frac{3\dot{a}}{a}\partial _t
-\frac{1}{a^2}\triangle\right)\Gamma \,, \label{einsteintt}\label{G^t_t}\\
\delta G^{x_i}_{\,\,t}&=&\partial _{x_i} \Bigg[\frac{2}{a^2}\partial _t\Psi
-\frac{1}{a^2}\left(\frac{\dot{b}}{b}+2\frac{\dot{a}}{a}\right)\Phi
-\frac{1}{a^2}\left(\frac{\dot{b}}{b}-\frac{\dot{a}}{a}+\partial _t\right)\Gamma
\Bigg] \,. \label{einsteinxit}
\end{eqnarray}
\end{widetext}

One can check that the equations of motion involving $W$ decouple from the other 
ones for the matter content (\ref{emt}) we consider. Since $W$ would appear only 
squared in a four dimensional effective theory, we will not need to compute it at 
all \footnote{That the equations of motion for $W$ decouple from the other ones is a 
direct consequence of the homogeneity in the $y$-direction.}.

To write down the perturbed Einstein equations
\footnote{Note that fluctuations in SGC (before dilaton stabilization) were 
considered in \cite{Scott2,Scott3}.}, we also need the perturbed 
energy-momentum tensor  $\delta T^A_{\,\,B}$. It will include the thermal bath of 
radiation (denoted by the subscript $r$), and the stringy matter sources denoted by 
a tilde.  To be specific, we have
\begin{eqnarray}
\nonumber (\delta ^{(r)}T^{A}_{\,\,B})&=&
\left(
\begin{array}{ccc}
\delta \rho_{(r)}&-(\rho_{(r)} +p_{(r)})V_{,i}&0\\
(\rho_{(r)} +p_{(r)})V_{,i}&\delta p_{(r)} \delta{^i_j}&0\\
0&0&0
\end{array}
\right)\,,
\end{eqnarray}
(where $\rho_{(r)}$ and $p_{(r)}$ are the radiation energy density and pressure,
respectively, and $V$ is the radiation three velocity potential) and
\begin{eqnarray}
\nonumber (\delta \tilde{T}^{A}_{\,\,B})&=&
\left(
\begin{array}{ccc}
\delta \tilde{\rho}&-(\rho_{(r)} +p_{(r)})\tilde{V}_{,i}&0\\
(\rho_{(r)} +p_{(r)})\tilde{V}_{,i}&\delta \tilde{p} \delta{^i_j}&0\\
0&0& \delta \tilde{r}
\end{array}
\right)\, ,
\end{eqnarray}
where ${\tilde{\rho}}$ and ${\tilde{p}}$ are the string gas energy 
density and pressure, respectively, and ${\tilde{V}}$ is the string gas three 
velocity potential.
Note that anisotropic stress does not feature in our setup (see appendix \ref{appendix}), but we will keep ${\tilde{V}}$ around for the time being. We will also focus on 
adiabatic perturbations of the radiation fluid only, that is 
$\delta p_{(r)}= \delta \rho_{(r)}/3$, even though two ideal fluids are present so 
that iso-curvature perturbations could arise. The reason for neglecting those is 
simplicity. The $x_i-x_j \,\, (i \neq j)$ Einstein equations yield immediately
\begin{eqnarray} 
\Gamma=\Psi +\Phi\,,
\end{eqnarray}
and, after introducing
\begin{eqnarray}
\xi&:=&\Psi-\Phi\\
N&:=&\frac{\delta\mu_0}{\mu_0} \, ,
\end{eqnarray}
we infer from (\ref{tt})-(\ref{aa}), that the perturbed energy momentum tensor is of
the form
\begin{eqnarray}
\delta \tilde{\rho}&=&(N+\Gamma+2\xi)\tilde{\rho}\,,\\
\delta\tilde{p}&=&(N+\Gamma+2\xi)\tilde{p}\,,\\
\delta\tilde{r}&=&(N+\xi)\tilde{r}-2\Gamma\left(\frac{1}{\tilde{b}^2}+\tilde{b}^2\right)3\tilde{p}\frac{a^2}{q^2\alpha^{\prime}}\,,\label{deltar}
\end{eqnarray}
where the background quantities are given by
\begin{eqnarray}
\tilde{\rho}&=&\frac{\mu_0}{\alpha^\prime}\frac{1}{a^3\tilde{b}}\sqrt{\frac{q^2\alpha^{\prime}}{a^2}+\left(\frac{1}{\tilde{b}}-\tilde{b}\right)^2}\,,\\
\tilde{p}&=&\frac{\mu_0}{\alpha^\prime}\frac{1}{a^3\tilde{b}}\frac{\frac{q^2\alpha^{\prime}}{3a^2}}{\sqrt{\frac{q^2\alpha^{\prime}}{a^2}+\left(\frac{1}{\tilde{b}}-\tilde{b}\right)^2}}\,,\\
\tilde{r}&=&\frac{\mu_0}{\alpha^\prime}\frac{1}{a^3\tilde{b}}\frac{\frac{1}{\tilde{b}^2}-\tilde{b}^2}{\sqrt{\frac{q^2\alpha^{\prime}}{a^2}+\left(\frac{1}{\tilde{b}}-\tilde{b}\right)^2}}\,.
\end{eqnarray}
In performing this calculation, we started with the
source action for this string gas \cite{sp1}, and obtained the
perturbed energy momentum tensor through direct calculation. The string
gas energy momentum tensor only depends on the metric, the number density
of strings $\mu_0$, and the center of mass momentum of these strings. However,
 the latter quantity is not perturbed, because for long
wavelength perturbations (compared to the string scale), which we are
restricted to if we are to remain within the limits of low energy
effective theory, the center of mass motion of a string propagating on a
perturbed spacetime is unaffected to first order. Hence the only quantities
left to perturb are the metric dependencies, and the number density
of the string gas.

Next, we can write the perturbed zero component of the conservation equation $\nabla ^{A}\tilde{T}_{AB}=0$ 
for the stringy matter as
\begin{eqnarray}
\label{eomNagain}0&=&\delta\dot{\tilde{\rho}}+3H(\delta\tilde{\rho}+\delta\tilde{p})+\mathcal{H}(\delta\tilde{\rho}+\delta\tilde{r})\\
\nonumber &&-\frac{3}{2}\dot{\xi}(\tilde{\rho}+\tilde{p})-\frac{1}{2}\dot{\Gamma}(\tilde{\rho}-3\tilde{p}+2\tilde{r})+\frac{\triangle \tilde{V}}{a^2}(\tilde{\rho}+\tilde{p})\,.
\end{eqnarray}
For simplicity we will set $\tilde{V}=0$ from now on, that is we neglect the scalar 
velocity potential of the string gas, consistent with the arguments of appendix \ref{appendix}. Combining the diagonal Einstein equations yields
\begin{eqnarray}
\nonumber &&\frac{1}{3M_5^3}\left(2\delta\tilde{r}+\delta\tilde{\rho}-3\delta\tilde{p}\right)=\ddot{\Gamma}+\dot{\Gamma}(3H+\mathcal{H})-2\mathcal{H}\dot{\xi}-\frac{\triangle}{a^2}\Gamma\\
&&\hspace{2cm} +(\Gamma -\xi)(3H\mathcal{H}+\mathcal{H}^2+\dot{\mathcal{H}})\,,\\
\label{eomxiagain} &&\frac{1}{3M_5^3}\left(3\delta\tilde{p}-\delta\tilde{\rho}\right)=\ddot{\xi}+\dot{\xi}(2\mathcal{H}+5H)-\mathcal{H}\dot{\Gamma}-\frac{\triangle}{3a^2}\xi\\
\nonumber&&\hspace{2cm}+(\xi-\Gamma)(4H^2+3H\mathcal{H}+\mathcal{H}^2+\dot{\mathcal{H}}+2\dot{H})\,.
\end{eqnarray}

The last three equations (\ref{eomNagain})-(\ref{eomxiagain}) are the dynamical ones 
for the perturbation variables $\Gamma$, $\xi$ and $N$. The ($t-t$) equation gives the 
radiation fluid perturbation $\delta\rho_{(r)}$ and the ($x_i-t$) equation gives the 
scalar velocity potential $V$ of the radiation fluid in terms of the other variables. 

These equations simplify if we write them in terms of $\eta$ defined in (\ref{defeta}) 
and make use of the background equation $a=a_0\eta$. Introducing 
$h:=\tilde{b}^{\prime}/\tilde{b}$ we get
\begin{eqnarray}
\nonumber && \Gamma_k^{\prime\prime}+\Gamma_k^{\prime}\left(h+\frac{2}{\eta}\right)-\xi_k^{\prime}2h+(\Gamma_k-\xi_k)\left[\frac{2h}{\eta}+h^2+h^{\prime}\right] \label{eomgamma}\\
&&\hspace{1cm}+\frac{k^{*2}}{4}\Gamma_k=\eta^2C\left(2\delta r^{*}+\delta\rho^{*}-3\delta p^{*}\right)\,,\\
\nonumber && \xi_k^{\prime\prime}+\xi_k^{\prime}2\left(h+\frac{2}{\eta}\right)-\Gamma_k^{\prime}h+(\xi_k-\Gamma_k)\left[\frac{2h}{\eta}+h^2+h^{\prime}\right]\\
&&\hspace{1cm}+\frac{k^{*2}}{12}\xi_k=\eta^2C\left(3\delta p^{*}-\delta\rho^{*}\right)\,,\\
&&\nonumber \delta\rho^{*\prime}+\frac{3}{\eta}(\delta\rho^{*}+\delta p^{*})+h(\delta\rho^{*}+\delta r^{*})-\frac{3}{2}\xi_k^{\prime}(\rho^{*}+p^{*})\\
&&\hspace{1cm}-\frac{1}{2}\Gamma_k^{\prime}(\rho^{*}-3p^{*}+2r^{*})=0\,.\label{eomrhoeta}
\end{eqnarray}
Here, we made the transition to Fourier space and we performed the rescaling 
$a^{*}:=a/a_0$, $k^{*}:=kt_0/a_0$. All other starred quantities are defined via 
\begin{eqnarray}
f^{*}:=f_ka_0^3\frac{\alpha^{\prime}}{\mu_0}\,.
\end{eqnarray}  
The two dimensionless constants left are
\begin{eqnarray}
C&:=&\frac{t_0^2}{M_5^3}\frac{1}{12a_0^3}\frac{\mu_0}{\alpha^{\prime}}\,,\\
q^{*}&:=&\frac{q\sqrt{\alpha^\prime}}{a_0}\,,\label{qstar}
\end{eqnarray}
which are not independent but related via 
\begin{eqnarray}
\frac{2^5C}{q^{*}}=1\,.
\end{eqnarray}

\subsection{Analytic late time solution}

After $h=h^{\prime}=0$ got approached (that is after the extra dimension got 
stabilized or, in the language of the ekpyrotic/cyclic scenario, the two approaching 
branes came to a halt), we can simplify the equations of the previous section to
\begin{eqnarray}
\Gamma_k^{\prime\prime}+\frac{2}{\eta}\Gamma_k^{\prime}+\frac{k^{*2}+\tilde{k}^2}{4}\Gamma_k&=&0\,,\\
\xi_k^{\prime\prime}+\frac{4}{\eta}\xi_k^{\prime}+\frac{k^{*2}}{12}\xi_k&=&0\,,
\end{eqnarray}
where  we introduced the constant
\begin{eqnarray}
\tilde{k}^2:=\frac{2^5C}{q^{*}}=1\,.
\end{eqnarray}
We note that the source term for $\xi_k$ vanishes because of the background 
equation of state for the gas of massless string modes. Also, the equations for $\xi_k$ and $\Gamma_k$ decouple at late times, just as 
they should do. The solutions are given by
\begin{eqnarray}
\Gamma_k&=&\frac{1}{\eta}\left(C_1\cos(\omega_{\Gamma}\eta)+C_2\sin(\omega_{\Gamma}\eta)\right)\,,\label{apprxGamma}\\
\xi_k&=&\frac{1}{\eta^3}\Big(C_3(k^{*}\eta\cos(\omega_\xi\eta)-2\sqrt{3}\sin(\omega_\xi\eta))\label{apprxxi}\\
\nonumber &&\hspace{0.6cm}+C_4(k^*\eta \sin(\omega_\xi\eta)+2\sqrt{3}\cos(\omega_\xi\eta))\Big)\,,
\end{eqnarray}
with $C_i$ constant and
\begin{eqnarray}
\omega_{\Gamma}^2&:=&\frac{k^{*2}+\tilde{k}^2}{4}\,,\\
\omega_{\xi}^2&:=&\frac{3k^{*2}}{36}\,.
\end{eqnarray}
We would like to evaluate the spectrum 
\begin{eqnarray}
\mathcal{P}_k:=k^3\Phi_k^2\propto k^{n_s-1}
\end{eqnarray}
when a long wavelength mode enters the Hubble radius again  at $k^{*}\eta_r=2$. If we 
Taylor expand (\ref{apprxxi}) for small $k^*$ we see that there is an approximately 
constant mode present for $\xi_k$. On the other hand, $\Gamma_k$ from 
(\ref{apprxGamma}) is oscillating and decaying $\propto 1/\eta$. Therefore, whatever 
the spectrum for $\xi_k$ was at the initial time, it should persist till re-entry and 
determine the spectrum for $\Phi_k$ because, by neglecting $\Gamma_k$, we have 
$\Phi_k\approx-\Psi_k\approx-\xi_k/2$. This is the same result one would conclude 
in a simple four dimensional universe, dominated by radiation.
To be specific, we can approximate
\begin{eqnarray}
\left|\Phi_k(\eta_r)\right|&\approx&\left|-\frac{\xi_k(\eta_i)}{2}+\frac{\Gamma_k(\eta_i)}{2\eta_r/\eta_i}\frac{\cos(\tilde{k}\eta_r/2+\beta)}{\cos(\tilde{k}\eta_i/2+\beta)}\right|\,, \label{analyticphi}\\
&\approx&\left|-\frac{\xi_k(\eta_i)}{2} \right| \,,
\end{eqnarray}
where $\beta$ is some irrelevant phase. Of course this holds true only, if the 
transient stabilizing epoch leaves no strong imprints onto the spectrum; hence, we 
will examine this crucial issue in the next section. 

Before we do that, let us have a quick look at the physics after a mode entered the 
Hubble radius: $\xi_k$ will start to oscillate with a decaying amplitude 
$\propto1/\eta^2$. If one waits long enough, $\Gamma_k$ will catch up and become 
larger than $\xi_k$. Now $\Phi_k\approx\Psi_k\approx\Gamma_k/2$ and thus $\Gamma_k$ 
will determine the spectrum, because for $\eta\gg\eta_r$
\begin{eqnarray}
\left|\Phi_k(\eta)\right|\approx\left|\frac{\Gamma_k(\eta_i)}{2\eta/\eta_i}\frac{\cos(\tilde{k}\eta/2+\beta)}{\cos(\tilde{k}\eta_i/2+\beta)}\right|\,.
\end{eqnarray}
There is of course a transition region where neither approximation holds. Nevertheless, 
this only occurs after Hubble radius crossing and shall not concern us in the following.

\subsection{Numerical solution and contact with late times \label{num}}

The analytic solution of the previous section is valid after $\tilde{b}$ got 
stabilized (or the branes came to a halt). However, the main question we would 
like to address is the following: what effect on the perturbations, if any, does 
the process of trapping the modulus $\tilde{b}$ have? In order to address this 
question we have to solve (\ref{eomgamma})-(\ref{eomrhoeta}) numerically. As we 
shall see, the late time solution is approached rapidly and long wavelength 
perturbations are not affected in any significant way. Hence, the naive intuition 
that the value of $\Phi_k$ for modes outside the Hubble radius stay frozen is 
indeed valid, with important 
consequences for cosmological model building, be it in the framework of the 
ekpyrotic/cyclic scenario, or within string gas cosmology (SGC) -- we will focus 
on concrete models in the next section.

\begin{figure}[tb]
  \includegraphics[width=\columnwidth]{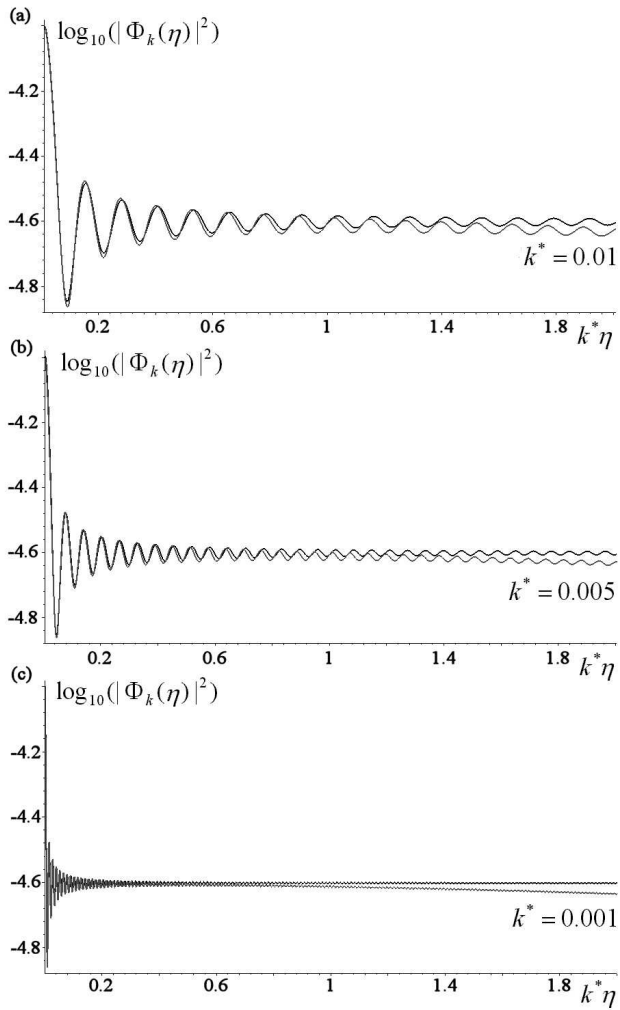}
   \caption{\label{fig:phi} $\log(|\Phi_k^2)|$ is plotted for different values of 
$k^*$, with the initial conditions given in section \ref{num}. Black: analytic 
solution of (\ref{analyticphi}); Grey (bending curve): numerical solution of 
(\ref{eomgamma})-(\ref{eomrhoeta}). }
\end{figure}

\begin{figure}[tb]
  \includegraphics[width=\columnwidth]{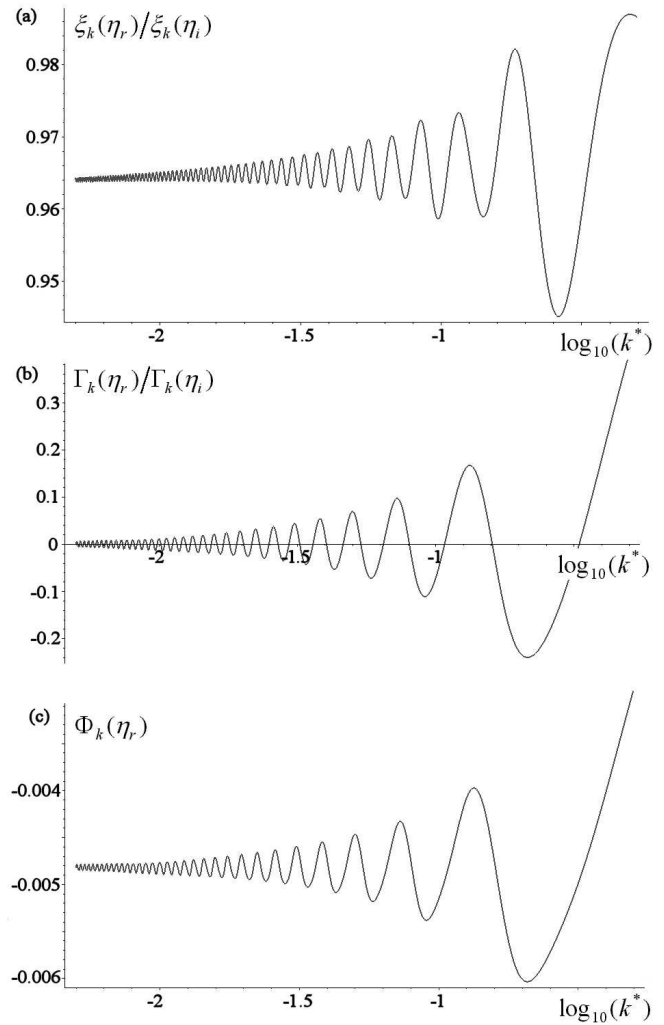}
   \caption{\label{fig:spectrum} The spectrum of (a) $\xi_k$, (b) $\Gamma_k$ and 
(c) $\Phi_k=(\Gamma_k-\xi_k)/2$ is evaluated at Hubble radius crossing $\eta_r=2/k^*$, 
with the initial conditions given in section \ref{num}. If the long wavelength 
regime $k^*\ll1$ is approached, all oscillations are damped away such that only 
the constant mode of $\xi_k$ survives, as expected.}
\end{figure}

\begin{figure}[tb]
  \includegraphics[width=\columnwidth]{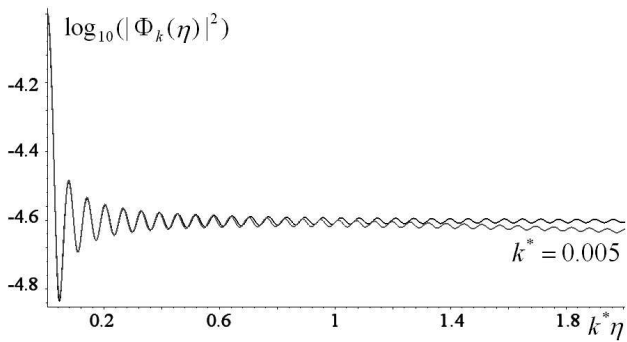}
   \caption{\label{fig:phi2_b} $\log(|\Phi_k^2)|$ is plotted for  
$k^*=0.005$. This is the same plot as in Fig.\ref{fig:phi} (b), but with different initial conditions for $\tilde{b}$: $\tilde{b}(\eta_i)=1$ and
$\tilde{b}^\prime(\eta_i)=0.1$ are used instead of (\ref{63}) and (\ref{64}). Note that there is no discernable difference between this plot and Fig.\ref{fig:phi} (b).}
\end{figure}

Firstly, let us specify constants and initial conditions: the only free parameter 
in our setup is $q^{*}$ defined in (\ref{qstar}), describing the initial momentum 
of the string gas in the three large dimensions. Being a dimensionless parameter, 
we choose the most natural value
\begin{eqnarray}
q^{*}:=1\,.
\end{eqnarray}
Our results are not very sensitive to its exact value \footnote{Decreasing $q$ 
induces a phase shift in $\Phi_k$ and an increase in $N_k$, while an increase in $q$ 
leads to a decrease in $N_k$ and no phase shift; however, the metric degrees of 
freedom decouple form $N_k$ quickly and $N_k$ itself is of no interest to us.}. We 
start close to the self-dual radius with some arbitrary velocity (once again, its 
exact value does not affect our results - see Fig.\ref{fig:phi2_b}), that is
\begin{eqnarray}
\tilde{b}(\eta_i)&=&1.1\,, \label{63}\\
\tilde{b}^\prime(\eta_i)&=&-0.1\,.\label{64}
\end{eqnarray} 

Next, we have to specify initial conditions for $\xi_k, \Gamma_k$ and $N_k$ at 
$\eta_i=1$ \footnote{We start at $\eta_i=1$, because this is the characteristic 
time scale of the background oscillations. This means, no oscillation could occur 
before $\eta_i$.}. Since we did not compute how the universe came close to the 
self-dual radius (e.g. within the ekpyrotic/cyclic setup or a SGC setup), we have no 
way of deriving those. However, we can make an educated guess: if the universe 
underwent some phase of inflation in the three dimension before (we will make this 
point more concrete in the next section), one should expect similar values for all 
metric perturbations; hence we will set 
\begin{eqnarray}
\xi_k(\eta_i)=\Gamma_k(\eta_i)=:\epsilon_k\,,
\end{eqnarray}
with some small $\epsilon_k$. For simplicity,  we will use $\epsilon_k=0.01$ for all 
$k$.  Furthermore, since the long wavelength modes we are interested in should be 
frozen once they cross the Hubble radius, we set
\begin{eqnarray} 
\xi_k^{\prime}(\eta_i)=\Gamma_k^{\prime}(\eta_i)=0\,.
\end{eqnarray}
The initial string density perturbation $N_k(\eta_i)$ will be set to zero, because 
these massless modes just got produced. Naturally, $N_k$ will get sourced by 
$\Gamma_k$ and $\xi_k$. No conclusion in the following is sensitive to the chosen initial 
conditions, hence we chose the most simple ones above (other ones were of course also tested).

The only thing left to specify is the wave-number $k^{*}$: We are interested in long 
wavelength modes, that is modes outside the Hubble radius with small $k^*$. In addition, 
we shall require $k^{*}<\tilde{k}$, so that we can compare our results with 
(\ref{analyticphi}). With all initial conditions specified, we can use the analytic solution
(\ref{analyticb}) to solve (\ref{eomgamma})-(\ref{eomrhoeta}) numerically.

\begin{figure}[tb]
  \includegraphics[width=\columnwidth]{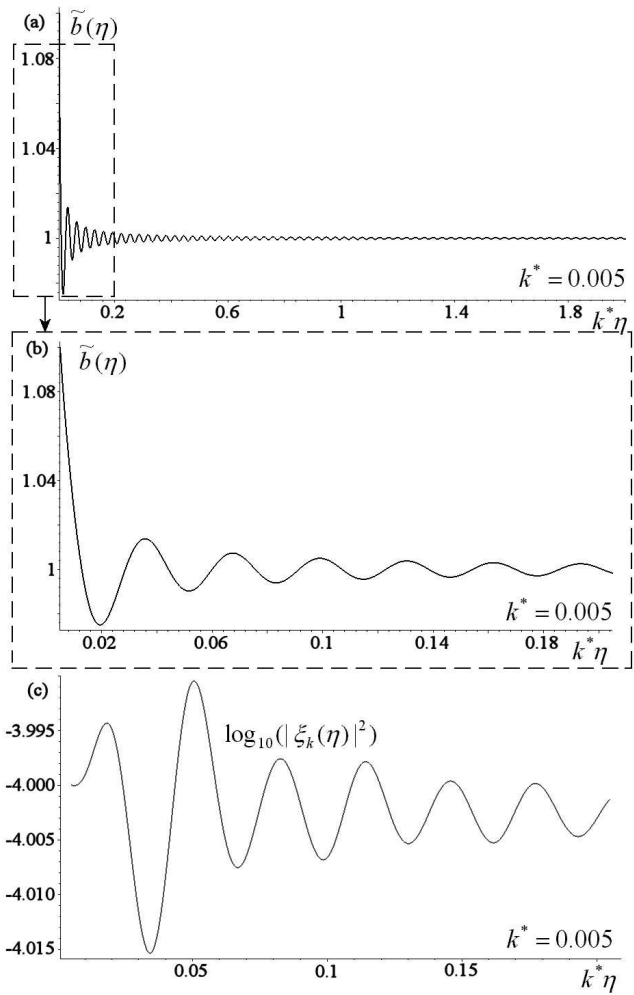}
   \caption{\label{fig:b} (a) and (b): $\tilde{b}$ over $k^{*}\eta$ is plotted for  
$k^*=0.005$, with the initial conditions given in section \ref{num}. (c): $\xi_k$ is 
plotted over the same time range as $b$; Note how the oscillations in the background 
scale factor source transient oscillations in $\xi_k$.}
\end{figure}

\begin{figure}[tb]
  \includegraphics[width=\columnwidth]{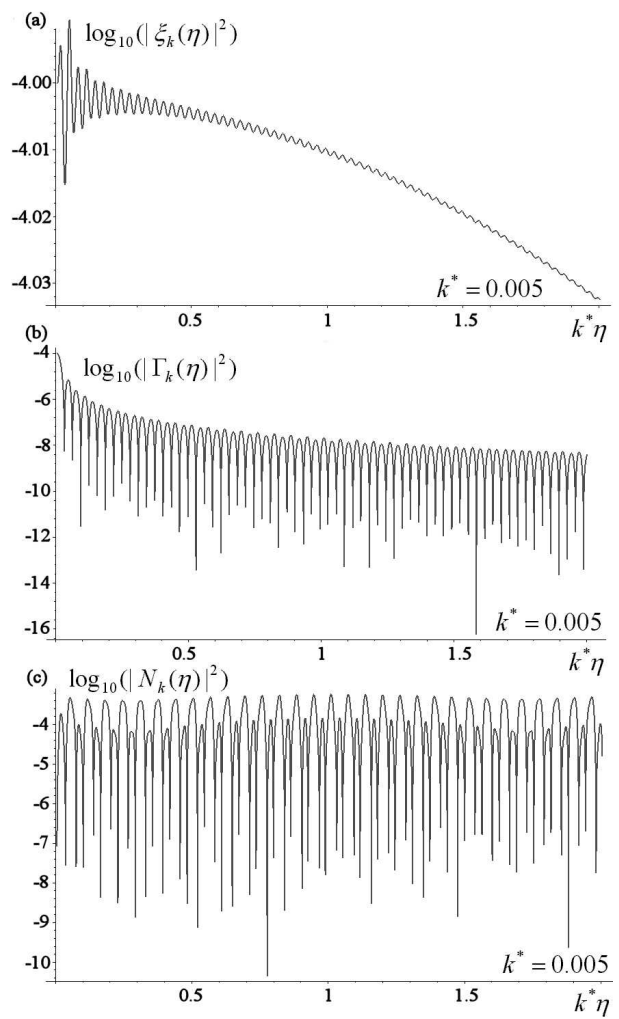}
   \caption{\label{fig:fields} The perturbation variables $\xi_k$, $\Gamma_k$ and 
$N_k$ plotted for  $k^*=0.005$, with the initial conditions given in section \ref{num}.}
\end{figure}

Let us first compare $\Phi_k(\eta)=(\Gamma_k(\eta)-\xi_k(\eta))/2$ for different 
values of $k^*$, Fig.~\ref{fig:phi} (a)-(c): There $\Phi_k$ is plotted both 
numerically (grey) and analytic (black). We see that the analytic late time solution 
is approached pretty fast -- in fact, the small difference at the beginning is not 
visible in this plot, but only if $\xi_k$ is plotted alone as in Fig.~\ref{fig:fields} 
(a). The visible decaying oscillation of $\Phi_k$ is the decaying mode of $\Gamma_k$ 
form (\ref{apprxGamma}) with frequency $\omega_\Gamma$, which is plotted in 
Fig.~\ref{fig:fields} (b) (compare with Fig.~\ref{fig:phi} (b)). In addition, there 
are strongly damped oscillations on top of the constant mode of $\xi_k$, that carry 
the same frequency as $\tilde{b}(\eta)$, Fig.~\ref{fig:b}. As mentioned before, these 
are not visible in Fig.~\ref{fig:phi} but in Fig.~\ref{fig:fields} (a) where $\xi_k$ 
is plotted alone. These oscillations are the impact of trapping the modulus 
$\tilde{b}$, an impact that can safely be ignored in the long wavelength regime: 
if one decreases $k^{*}$, all of these effects become less pronounced.

An other feature visible in Fig.~\ref{fig:phi} is a bending of the lower curve in 
comparison to the analytic solution (\ref{analyticphi}). This bending is the beginning 
of an oscillation of $\xi_k$ and hence an expected feature, since the constant mode is 
actually the beginning of an oscillation with a very small frequency due to 
$\omega_\xi$. All features are not sensitive to the initial values we choose for 
$\Gamma_k$ and $\xi_k$.

Last but not least, $N_k$ is plotted in Fig.~\ref{fig:fields} (c): it gets sourced 
quickly by $\Gamma_k$ and $\xi_k$ and oscillates with a nearly constant amplitude 
thereafter, with all frequencies entering. However, there is no mentionable 
back-reaction of $N_k$ on the metric perturbations. Hence, one could give $N_k$ 
a non zero initial value without changing the evolution of $\Gamma_k$ and $\xi_k$ 
in any significant way.

\section{Consequences for ekpyrotic/cyclic models and SGC}   

In the end, we are interested in the spectrum of $\Phi_k$. So far, there seems to be 
little to no impact of the transient stabilizing era of $b$ onto $\Gamma_k$ and 
$\xi_k$. This translates directly to the spectrum, where $\Phi_k$ should approach a 
horizontal line for small $k^*$. This is indeed the case, as can be seen in 
Fig.~\ref{fig:spectrum} (c), where the spectrum is evaluated at horizon crossing: 
the oscillations that are present for relatively large $k^*$ get damped once the 
long wavelength regime is approached. This is expected, because long wavelength modes 
enter the Hubble radius later and henceforth, the decaying modes responsible for the 
oscillations in the spectrum get damped more. The frequency in the spectrum of $\Phi_k$ 
is proportional to $1/k^*$, in accordance with $\eta_r\sim 1/k^*$ and equation 
(\ref{analyticphi}). 

Note that no shift of the spectral index is induced, leaving the 
overall index unaltered. The oscillations in $\xi_k$ in Fig.~\ref{fig:spectrum}(a) are 
the main impact of the transient era of a dynamical $b$. They are clearly negligible, 
because the tiny amplitude gets even smaller with decreasing $k^*$.

To summarize, the main conclusion of our numerical study is that long wavelength 
metric perturbations quickly approach their asymptotic solution, leaving no trace 
of the many bounces $b$ experiences on its way to stabilization. More specifically,
we have shown that the initial spectrum of the non-decaying mode of $\xi_k$ 
is preserved, i.e. equals the spectrum of $\xi_k$ at late times. At late
times, the spectrum of $\xi_k$ equals the spectrum of the Bardeen potential
$\Phi_k$, the potential which is relevant for late time observations. Speaking
more loosely, we have shown that the initial pre-bounce spectrum of bulk
perturbations is transferred to the late time cosmological perturbations without a
change in the spectral index. 

We saw in the previous section that there is no significant effect of trapping the 
modulus $\tilde{b}$ on the spectrum of long wavelength perturbations in the
sense that the final spectrum of the Bardeen potential $\Phi$ equals the initial
value of the non-decaying mode of the bulk perturbation $\xi$. This comes about 
since the value of $\xi$ for small $k^{*}$ modes which have a wavelength
larger than the Hubble radius remains frozen during the 
phase of the trapping of $\tilde{b}$, and since the final spectrum of $\xi$
coincides with the spectrum of $\Phi$. 

Even though our result agrees with a ``naive'' intuition coming from
the analysis of fluctuations in expanding four space-time dimensions, it 
could be viewed as an unexpected result: 
The ``naive'' intuition turned out to be wrong in the case of bouncing cosmologies 
in four space-time dimensions, for example in pre-big-bang models
where it was shown \cite{BGGMV} that the growing mode of $\Phi$ in the
contracting phase couples almost exclusively to the decaying mode of $\Phi$
in the expanding phase, leading to the result that there is a large change in
the spectral index of the dominant mode between the contracting and the expanding
phase. Other analyses performed within the context of four-dimensional
general relativity 
(see \cite{Bozza:2005wn,Bozza:2005xs,Battefeld:2005cj,Geshnizjani:2005hc} and references 
therein) yielded a similar result. It is now also generally accepted that
the four space-time dimensional toy models which were proposed to describe the
ekpyrotic/cyclic models have the same feature, namely that the dominant mode of
$\Phi$ in the contracting phase matches predominantly to the decaying mode of $\Phi$
in the expanding phase (see e.g. \cite{FB}). Our analysis, however, confirms
the analysis of \cite{Tolley}, which showed, in the context of a singular bounce,
that in the case of a bounce of boundary branes in five space-time dimensions
the spectrum of $\Phi$ could be preserved. 

Our result is an important step in constructing a viable alternative to 
standard scalar field-driven inflationary models (next to the remarkable but highly 
nontrivial KKLMMT construction, \cite{Kachru:2003sx} and follow-up papers)
in the context of cosmologies with extra dimensions. We will 
now outline two possible proposals which seem promising to us in that regard:

Firstly, one could envisage a modification of the ekpyrotic/cyclic scenario. 
In the ekpyrotic proposal \cite{Khoury:2001wf}, a test brane moves slowly through 
the bulk towards the boundary brane of an orbifold (the Horava-Witten 
\cite{HW} setup). In the cyclic scenario \cite{st1}, it is the two boundary branes 
that approach each other. In either case, a scale-invariant spectrum of perturbations 
can be generated during the contracting phase given a suitable potential for the 
modulus field which describes the inter-brane distance \cite{KOST2}. 
During the collision of the branes (a singular event in the ekpyrotic/cyclic scenarios) 
a hot universe is supposed to emerge on the boundary brane we live on today. 
Our background construction can be viewed as a regularized version of the
ekpyrotic/cyclic scenarios (ekpyrotic in the context of a single bounce, cyclic
in the case we consider the evolution until $\tilde b$ has stabilized). Our
work shows that in the context of such a regularized scenario, an initial
scale-invariant spectrum could pass through the bounces and thus survive from the
initial contracting phase to the final phase of expansion of our three spatial
dimensions (this scenario will be developed further in a follow-up paper).
In our scenario, the branes do not actually hit each other, but come to a halt as 
a consequence of the appearance of new massless modes that get produced 
explosively at a certain brane-separation (that is at a certain value of $b$). Our 
results show that such dynamics does not spoil a scale invariant spectrum 
generated during the initial brane movement. In this framework, one can still 
reheat the Universe, since the 
``kinetic energy'' of the branes will get transferred to the stabilizing massless 
modes and other light modes which act like radiation on the brane (corresponding to the 
radiation bath that we already included in our setup).   

Secondly, an incorporation of inflation into brane/string gas cosmology will rely on 
the stabilization mechanism described in this article: if any mechanism of inflating 
three dimensions is found, the internal dimensions will most likely have to deflate 
in some way. For example, one could employ the idea of anisotropic inflation (studied
in the context of vacuum solutions of higher dimensional general relativity in  
\cite{KKinflation} and \cite{Levin:1994yw}). Once the internal dimensions get close to 
the self-dual radius, certain string modes become massless and get produced explosively 
along with a radiation bath, in close analogy to reheating after standard scalar 
field-driven inflation. These modes will then stabilize the internal dimensions, 
curing the graceful exit problem of anisotropic inflation, while the spectrum of 
fluctuations produced during inflation in the metric degrees of freedom remains 
unaltered. A concrete realization of this proposal is in preparation by the authors 
of this article.

\section{Conclusions}

In this report, we studied how radion, matter and metric fluctuations interact in a 
universe that exhibits a transient stabilizing epoch of its extra dimension, which 
could also be viewed as a series of bounces of the extra dimension. We were primarily 
interested in the imprints of this epoch on the spectrum of the Bardeen potentials, 
motivated by the hope of combining an earlier phase, generating a scale invariant 
spectrum (e.g. in the framework of a modified ekpyrotic scenario, or via an 
incorporation of inflation within string gas cosmology), with a successful late time 
stabilizing mechanism, provided by a gas of massless string states. 

We found that the spectrum of long wavelength perturbations remains unaltered by this 
epoch, which has important consequences for various approaches to stringy models of 
the early universe. In particular, specific realizations of the ekpyrotic/cyclic 
scenario or an incorporation of inflation into string gas cosmology become possible -- 
we provided two proposals in this report, but many variations are viable.

\begin{acknowledgments}

We would like to thank Scott Watson for helpful comments and discussions, and also 
Andrei Linde, Diana Battefeld and Claudia De Rahm for comments on the draft. T.B. would like to thank the Physics 
Department of McGill University for hospitality and support. S.P. wishes to thank the 
theory group at Queen Mary University in London and the cosmology group at DAMTP in 
Cambridge for hospitality during the course of this work. This research is 
supported at McGill by an NSERC discovery grant, by funds from the Canada Research
Chair program, and by funds from McGill University. T.B. has been supported in part
by the US Department of Energy under contract DE-FG02-91ER40688, TASK A.

\end{acknowledgments}

\appendix
\section{Anisotropic stress \label{appendix}}

In this appendix, we will provide arguments as to why no anisotropic stress arises in the setup of  \cite{sp1}, which we examined at the perturbative level in this article. In the following $a,b,\dots$ denote string world-sheet coordinates.

First note that our unperturbed background is undergoing FRW expansion in the non-compact dimensions while the compact direction is stabilized at the self dual radius by a string gas. As a preliminary, we wish to show that long wavelength perturbations of the metric do not affect the motion of strings to first order in the perturbation variables. We begin with the Polyakov action for a closed string
\begin{equation}
\label{polyclosed}
S = -\frac{1}{4\pi \alpha^{\prime}} \int d^2 \sigma \partial^a X^A \partial_a X^B g_{A B}(X)\,,
\end{equation}
where we work in conformal gauge from the outset. Consider now perturbations around the background: $g_{AB} = g^0_{AB} + h_{AB}$. 
It was shown in the appendix of \cite{sp1} how one can inherit the string spectrum and the constraint algebra provided the derivatives of the background metric $g^0_{AB}$ (i.e. the Hubble factors) are several orders of magnitude smaller than the string energy scales. We want to make sure that we stay within this regime while studying the dynamical compactification of the extra dimension. Thus, when we consider the equations of motion for a closed string in this perturbed spacetime, we must neglect all derivatives of the unperturbed metric compared to the worldsheet derivatives of the string co-ordinates, as these will be of the order of the string scale whereas the metric derivatives are constrained to be much smaller \cite{sp1}. 

Consider the equation of motion for a closed string 
\begin{eqnarray*}
\partial_a \Bigl(\partial^a X^A g_{C A}(X) \Bigr) &=& \frac{1}{2}\partial_aX^A \partial^a X^B \partial_C g_{A B}(X)\,,
\end{eqnarray*}
from which follows
\begin{eqnarray*}
\frac{1}{2} \partial_a X^A \partial^a X^B \partial^C h_{A B} &=& \partial_a\partial^a X^C + h^C_A \partial_a\partial^a X^A \\
&&+ \partial_a X^A \partial^a X^B \partial_A h_{B D}g^{D C}_0\,.
\end{eqnarray*}
Now we expand $X^A$ as $X^A = X_0^A + \delta X^A$, where $X^A_0$ is the solution to the unperturbed equation of motion $\partial_a\partial^a X^A_0 = 0$, see appendix of \cite{sp1}. Our perturbed equation of motion then becomes
\begin{eqnarray}
 \frac{1}{2} \partial_a X_0^A \partial^a X_0^B \partial^C h_{A B}& = &\partial_a\partial^a \delta X^C \label{perteom}\\
\nonumber &&+ \partial_a X_0^A \partial^a X_0^B \partial_A h_{B D}g^{D C}_0\,.
\end{eqnarray}
Notice that there are two vastly different scales in the equation above: the worldsheet derivatives and the spacetime derivatives. If we imagine expanding $h_{A B}$ in terms of fourier modes, the derivatives would bring down a factor of the 4-d wave-vector. Realizing that we need to stay within the domain of validity of our setup, namely that we cannot introduce fluctuations in the metric that would invalidate the conclusion in the appendix of \cite{sp1} concerning the string spectrum, we need these fluctuations to vanish on energy scales several orders of magnitude smaller than the string scale. 
This still allows for metric fluctuations that are considerably short distance in nature, but nevertheless above the string scale. With this in mind, we can neglect all the terms in (\ref{perteom}) that are multiplied by derivatives of the metric perturbation, since these will be orders of magnitude smaller than the other ones. Henceforth, we obtain the result
\begin{equation}
\label{pertmom}
\partial_a\partial^a \delta X^A = 0 \,,
\end{equation}    
that is, the string motion is unchanged by long wavelength perturbations. 

Ordinarily this would signal the end of any hopes to study perturbative physics; this is not the case for us, since we are primarily interested in perturbing around a background which consists of a gas of such strings.
As a result, if we consider the perturbed energy-momentum tensor for such a gas, it will contain no anisotropic stress. To see this, consider the off-diagonal spatial components which can be written as
\begin{equation}
\label{pertspat}
\delta T^i_{\,\,j} \propto  \langle p^i \delta p_j \rangle + \langle \delta p^i p_j \rangle\,,
\end{equation}
where $i\neq j$ and  $\langle ... \rangle$ indicates an ensemble (gas) average. These components vanish, since we just determined that for individual strings $\delta p^i = 0$, 
which means that there is no anisotropic stress to deal with in our setup.

\end{document}